\documentclass[12pt]{article}
\usepackage{graphicx}
\usepackage{amssymb}
\def\lsim{\mathrel{\raise.3ex\hbox{$<$\kern-.75em\lower1ex\hbox{$\sim$}}}}
\def\gsim{\mathrel{\raise.3ex\hbox{$>$\kern-.75em\lower1ex\hbox{$\sim$}}}}

\begin{document}

\title{Testing the LMA solution with solar neutrinos \\independently of solar models}
\author{V. Barger$^1$, D. Marfatia$^{2}$ and K. Whisnant$^3$\\[2ex]
\small\it $^1$Department of Physics, University of Wisconsin, Madison, WI 53706\\
\small\it $^2$Department of Physics and Astronomy, University of Kansas, Lawrence, KS 66045\\
\small\it $^3$Department of Physics, Iowa State University, Ames, IA 50011}

\date{}

\maketitle

\begin{abstract}
We perform a comparative study of
two methods of determining the survival 
probabilities of low, intermediate, and high energy solar neutrinos that
emphasizes the general agreement between 
the Large Mixing Angle (LMA)
solution and extant solar neutrino data.
The first analysis is oscillation parameter-independent and 
the second analysis involves
an approximate calculation of the survival probabilities in the three 
energy ranges that depends only on oscillation parameters. 
We show that future experiments like 
BOREXino, CLEAN, Heron, LENS and MOON, that measure $pp$ and $^7$Be
neutrinos, will facilitate a stringent test of the LMA solution
independently of the Standard Solar Model (SSM), without recourse to
earth-matter effects. Throughout, 
we describe the role of SSM assumptions on our results. If the LMA solution
passes the test without needing to be modified, it may be possible to
establish that $\theta_x$ is nonzero at more than $2\sigma$ 
assuming the SSM prediction for the $pp$ flux is correct.

\end{abstract}

\newpage

\section{Introduction}

The LMA solution to the solar neutrino problem
has emerged as the preferred explanation of solar neutrino 
data~\cite{Barger:2002iv,catch}. 
Reactor neutrino data from the KamLAND experiment~\cite{Eguchi:2002dm} 
have lent much confidence
in this solution~\cite{Barger:2002at}.
Nevertheless, it has long been recognized that the LMA 
solution is not in wholly satisfactory 
agreement with solar neutrino data~\cite{Barger:2002iv}.
Since the media traversed by neutrinos incident at KamLAND and at solar
neutrino experiments are considerably different, it is important to confirm
that the LMA solution with matter effects dictated by the MSW mechanism is
consistent with solar neutrino data. Recent work emphasizing our ignorance
of neutrino-matter interactions and suggestions on how solar data can help
illuminate the nature of these interactions can be found 
in Ref.~\cite{Fardon:2003eh}.
  
For the LMA solution, the Chlorine measurement~\cite{Cleveland:1998nv}, 
 with its non-negligible component of $^7$Be neutrinos, is
too low to be consistent with the SuperKamiokande~\cite{Fukuda:2002pe} 
(SuperK) and 
SNO~\cite{Ahmad:2002jz} measurements of the
$^8$B neutrinos. The Small Mixing Angle solution can accommodate a low
survival probability of $^7$Be neutrinos, but at the expense of a
highly non-uniform suppression of the $^8$B neutrinos relative to the
SSM~\cite{Bahcall:2004fg} spectrum, which is contradicted by SuperK
and SNO data. Exotic mechanisms have been proposed to account for the
suppression of the $^7$Be neutrinos, but these require that
either a sterile neutrino be added~\cite{smirnov} 
to the standard three-neutrino framework
or that neutrino masses vary with density~\cite{mavans}.

In this Letter we perform a model-independent analysis
of the latest
solar neutrino data with the flux normalizations and survival
probabilities of the low, intermediate, and high energy neutrinos all
treated as independent parameters; the possibility of a model-independent 
test of the SSM was emphasized early on in Ref.~\cite{hata}. 
We then determine which of these
parameters are calculable from current data and which require
additional input from the SSM. The results of this analysis are then
used to test the LMA predictions for neutrinos in each energy
range. We show that with the predictions of the latest SSM and
including the latest SNO salt data~\cite{salt} the agreement between the LMA
solution and the solar data remains relatively good, although there
are minor discrepancies between the oscillation probabilities of the
low- and intermediate-energy solar neutrinos which cannot be resolved
by extending the analysis to include the mixing of a third
neutrino. We demonstrate that future solar neutrino experiments that
would measure low and intermediate energy neutrinos and reduce the
uncertainties in the survival probabilities will provide a critical
test of the LMA solution.

\section{Current data} 

\subsection{Model independent analysis}

\subsubsection{Formalism}

We use the following notation: $R$ is the ratio of the measured rate to the
SSM prediction for a given experiment, $\beta$ is a
flux normalization relative to the SSM, and $P_L$, $P_I$ and $P_H$ are 
average survival probabilities of low energy ($pp$), intermediate
energy ($^7$Be, $pep$, $^{15}$O, $^{13}$N) and high energy ($^8$B, $hep$) 
neutrinos, respectively. 
For three active neutrinos and the recently 
updated SSM~\cite{Bahcall:2004fg}, the relative rates are given by
\begin{eqnarray}
R_{Ga} &=& 0.109 \beta_H P_H + 0.335 \beta_I P_I + 0.556 \beta_L P_L \,,
\label{eq:RGa}
\\
R_{Cl} &=& 0.803 \beta_H P_H + 0.197 \beta_I P_I \,,
\label{eq:RCl}
\\
R^{CC}_{SNO} &=& \beta_H P_H \,,
\label{eq:RSNOCC}
\\
R^{NC}_{SNO} &=& \beta_H \,.
\label{eq:RSNONC}
\end{eqnarray}
Here, CC and NC refer to charged-current and neutral-current measurements,
respectively. See Refs.~\cite{reexam,unknowns} for a description of 
the method. The present measurements of these quantities are given in
Table~\ref{tab:data}.  The elastic scattering rates at
SuperK and SNO are not included since they provide
redundant information with less precision than $R_{SNO}^{CC}$ and
$R_{SNO}^{NC}$, although in principle these data could also be included
to improve the accuracy.

\begin{table}[h]
\caption[]{Current solar neutrino measurements. \label{tab:data}}
\vskip0.1in
\centering\leavevmode
\begin{tabular}{ccl}
\hline
Measurement & value & source\\
\hline
$R_{Ga}$       & $0.54 \pm 0.03$ & SAGE/GALLEX/GNO~\cite{Abdurashitov:2002nt}\\
$R_{Cl}$       & $0.31 \pm 0.03$ & Homestake~\cite{Cleveland:1998nv}\\
$R_{SNO}^{CC}$ & $0.30 \pm 0.01$ & SNO D$_2$O phase, SNO salt phase~\cite{Ahmad:2002jz,salt}\\
$R_{SNO}^{NC}$ & $0.88 \pm 0.06$ & SNO D$_2$O phase, SNO salt phase~\cite{Ahmad:2002jz,salt}\\
\hline
\end{tabular}
\end{table}

\subsubsection{Without SSM constraints}

The quantities $\beta_L P_L$, $\beta_I P_I$ and $\beta_H P_H$ are
determined from the CC measurements $R_{Ga}$, $R_{Cl}$ and
$R^{CC}_{SNO}$. $R^{NC}_{SNO}$ determines $\beta_H$. Using the data in Table~\ref{tab:data}, we find
\begin{eqnarray}
\beta_L P_L &=& 0.69 \pm 0.11 \,,
\label{eq:BLPL}
\\
\beta_I P_I &=& 0.37 \pm 0.15 \,,
\label{eq:BIPI}
\\
\beta_H P_H &=& 0.30 \pm 0.01 \,,
\label{eq:BHPH}
\\
\beta_H &=& 0.88 \pm 0.06\,,
\label{eq:BH}
\end{eqnarray}
where the uncertainties are $1\sigma$. 
Then,
\begin{equation}
P_H = R_{SNO}^{CC}/R_{SNO}^{NC} = 0.34 \pm 0.03 \,.
\label{eq:PH}
\end{equation}
We note that $P_H$ would not be determined by $R_{SNO}^{CC}$ and
$R_{SNO}^{NC}$ alone if there were a fourth, sterile neutrino since the
expression for $R_{SNO}^{NC}$ would depend on the sterile content of
the oscillating neutrinos~\cite{unknowns}. 
At present there are no data that can
determine $\beta_L$ and $P_L$ (or $\beta_I$ and $P_I$) separately in a
completely model-independent way without imposing SSM constraints.

\subsubsection{With SSM constraints}

The $1\sigma$ fractional uncertainties of the solar
neutrino fluxes from the SSM~\cite{Bahcall:2004fg} are
\begin{eqnarray}
\delta(\beta_L^{SSM}) &=& 0.010 \,,
\label{eq:BL}
\\
\delta(\beta_I^{SSM}) &=& 0.122 \,,
\label{eq:BI}
\\
\delta(\beta_H^{SSM}) &=& 0.163 \,.
\label{eq:BH2}
\end{eqnarray}
The values of $P_L$ and $P_I$
can be determined from current data only if the SSM constraints are
used; from Eqs.~(\ref{eq:BLPL}, \ref{eq:BIPI}, \ref{eq:BL}) and
(\ref{eq:BI}) we deduce
\begin{eqnarray}
P_L &=& 0.69 \pm 0.11 {\rm~~with~SSM~flux} \,,
\label{eq:PL}
\\
P_I &=& 0.37 \pm 0.16 {\rm~~with~SSM~flux} \,.
\label{eq:PI}
\end{eqnarray}

If the SSM prediction for $\beta_H$ is used as an additional
constraint, then a combination of the SSM and $R_{SNO}^{NC}$ gives
\begin{equation}
\beta_H = 0.89 \pm 0.06 \,,
\label{eq:BH3}
\end{equation}
from which we deduce
\begin{equation}
P_H = 0.34 \pm 0.03 \,.
\label{eq:PH2}
\end{equation}

\subsection{LMA analysis}

\subsubsection{Formalism}

If the $\nu_e$ disappearance of solar neutrinos occurs via the
MSW mechanism~\cite{msw} with adiabatic propagation~\cite{adiabatic} 
(which is the situation for
the LMA solution), the solar neutrino oscillation probability in a
three-neutrino framework is given by the approximate formula~\cite{Kuo:1989qe}{\footnote{In our
notation, $\delta m^2_s$ is the solar mass-squared difference and $\theta_s$
and $\theta_x$ are the mixing angles conventionally denoted by $\theta_{12}$
and $\theta_{13}$, respectively~\cite{Barger:2003qi}.}},
\begin{equation}
P = c^4_{x}\left[ {1 + \cos2\theta_{s}^m \cos2\theta_{s}\over 2}\right]
+ s^4_{x} \,.
\label{eq:P}
\end{equation}
The above equation applies when $\theta_x$ is small, as is indicated by the
CHOOZ experiment~\cite{Apollonio:1999ae}; the 95\% C.~L. is
$\sin^2 \theta_x \lsim 0.05$ for the best-fit atmospheric mass-squared 
difference of 0.0021 eV$^2$ from the SuperK experiment~\cite{superkatm}. 
For the current range of LMA parameters preferred by data, earth-matter 
effects~\cite{baltz} are very small~\cite{Fukuda:2002pe,Ahmad:2002jz} and therefore we do not consider them here.
The quantity in the square brackets is the two-neutrino
oscillation probability and the factors involving $\theta_{x}$ are
the corrections due to mixing with the third neutrino. The angle
$\theta_{s}^m$ is the mixing angle in matter at the point of
origin of the neutrino, and is given by
\begin{equation}
\tan2\theta_{s}^m = {\sin2\theta_{s} \over \cos2\theta_{s} - \hat A} \,,
\label{eq:t^2}
\end{equation}
where
\begin{equation}
\hat A \equiv {2 \sqrt2 G_F N_e^0 E_\nu c_{x}^2 \over \delta m^2_{s}}
\simeq 1.9\times10^{-3} \left({E_\nu \over 1 {\rm~MeV}}\right)
\left({8\times10^{-5}{\rm~eV}^2\over\delta m^2_{s}}\right)
\left({N_e^0\over N_A/{\rm cm}^3}\right) c_{x}^2 \,.
\label{eq:A}
\end{equation}
The initial electron densities at the neutrino source are
approximately $N_e^0/(N_A/{\rm cm}^3) = 106$, 86 and 58 for the
high, intermediate and low energy neutrinos, for which the
corresponding average neutrino energies are 9.06, 0.862, and
0.325~MeV, respectively. The factor $c_{x}^2$ in $\hat A$ is a
three-neutrino correction to the effective electron number density~\cite{Kuo:1989qe}. In
the LMA analysis with three neutrinos, the variables that determine
the probabilities are therefore $\delta m^2_{s}$, $\theta_{s}$ and
$\theta_{x}$, instead of the oscillation parameter-independent 
probabilities $P_j$.

\subsubsection{Without SSM constraints}

\begin{figure}[ht]
\centering\leavevmode
\includegraphics[width=4in]{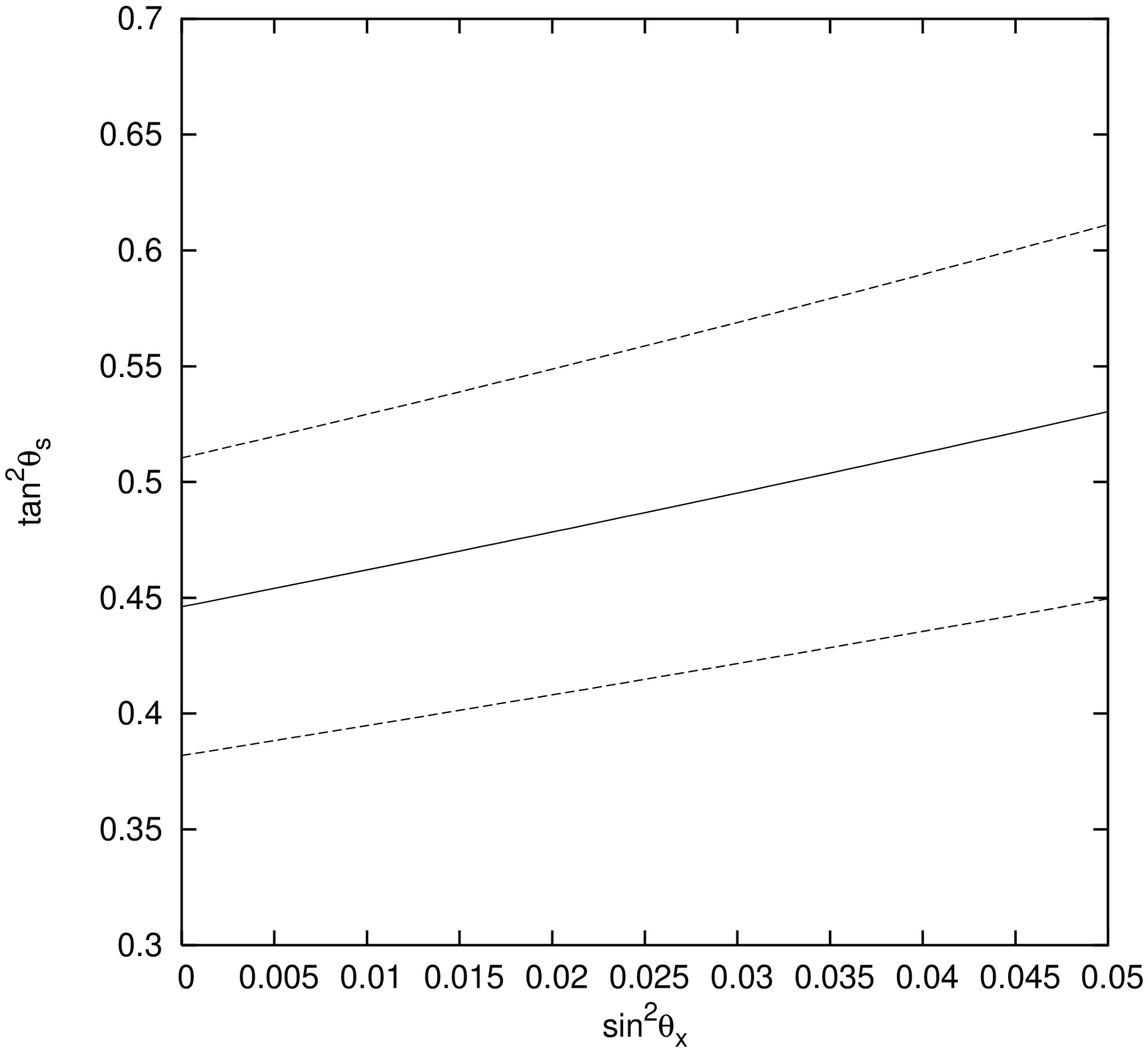}
\caption[]{Inferred values of $\tan^2\theta_{s}$ versus
$\sin^2\theta_{x}$ using the current SNO data and the $1\sigma$ range,
$\delta m^2_{s}= (7.9\pm 0.55)\times10^{-5}$~eV$^2$, from
KamLAND data alone. \label{fig:t2th12}}
\end{figure}

As discussed in the model-independent analysis, neutrino data alone
cannot presently determine $P_L$ or $P_I$, so these probabilities
cannot be used to constrain the LMA parameters. However, the SNO CC
and NC data can be used to determine $P_H$, and $\theta_{s}$ can then be
determined via Eqs.~(\ref{eq:P}--\ref{eq:A}) if
a range of $\delta m^2_{s}$ is used as input and we ignore
$\theta_{x}$. Taking the $1\sigma$ range, $\delta m^2_{s}= (7.9\pm0.55)\times10^{-5}$~eV$^2$, from KamLAND data~\cite{Eguchi:2002dm}, we find
\begin{equation}
\tan^2\theta_{s} = 0.45 \pm 0.06,
\label{thetas}
\end{equation}
in good agreement with recent global fits to $\theta_{s}$ that also
include day/night spectral shapes in the analysis~\cite{Balantekin:2004hi}. If
$\theta_{x}$ is allowed to be nonzero, $\theta_{s}$ can be
calculated as a function of $\theta_{x}$; see
Fig.~\ref{fig:t2th12}. An increasing $\theta_{s}$ can be compensated
by an increasing $\theta_{x}$.

\begin{figure}[ht]
\centering\leavevmode
\includegraphics[width=4in]{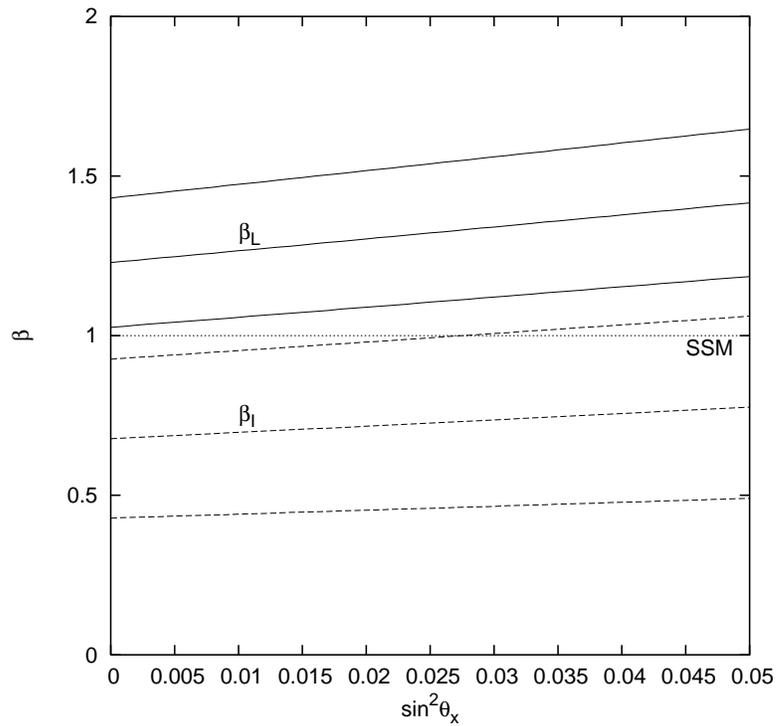}
\caption[]{Inferred values of $\beta_L$ (solid) and $\beta_I$ (dashed)
for $\delta m^2_{s}=(7.9\pm0.55)\times10^{-5}$~eV$^2$,
using the LMA predictions for $P_L$ and $P_I$ and Eqs.~(\ref{eq:BLPL},\ref{eq:BIPI}). In each case
the central values and $1\sigma$ error band is shown. The dotted
line is the SSM prediction. \label{fig:beta}}
\end{figure}

Since $\beta_L$ and $\beta_I$ are as yet undetermined by the data, we
can use the LMA prediction for $P_L$ and $P_I$ in conjunction with the
current best-fit values of $\beta_L P_L$ and $\beta_I P_I$ from
Eqs.~(\ref{eq:BLPL}) and (\ref{eq:BIPI}) to solve for $\beta_L$ and
$\beta_I$ (see Fig.~\ref{fig:beta}); the LMA predictions assume the 
$1\sigma$ ranges, 
$\delta m^2_{s}=(7.9\pm 0.55)\times10^{-5}$~eV$^2$ from KamLAND and  
$\tan^2\theta_{s}=0.45\pm 0.06$ from Eq.~(\ref{thetas}). 
We see that the flux
normalization of the low (intermediate) energy neutrinos needs to be
slightly higher (lower) in order for the LMA predictions to be completely
consistent with current data, although the present uncertainties are
sufficiently large that there is essentially no conflict.

\subsubsection{With SSM constraints}

\begin{figure}[ht]
\centering\leavevmode
\includegraphics[width=4in]{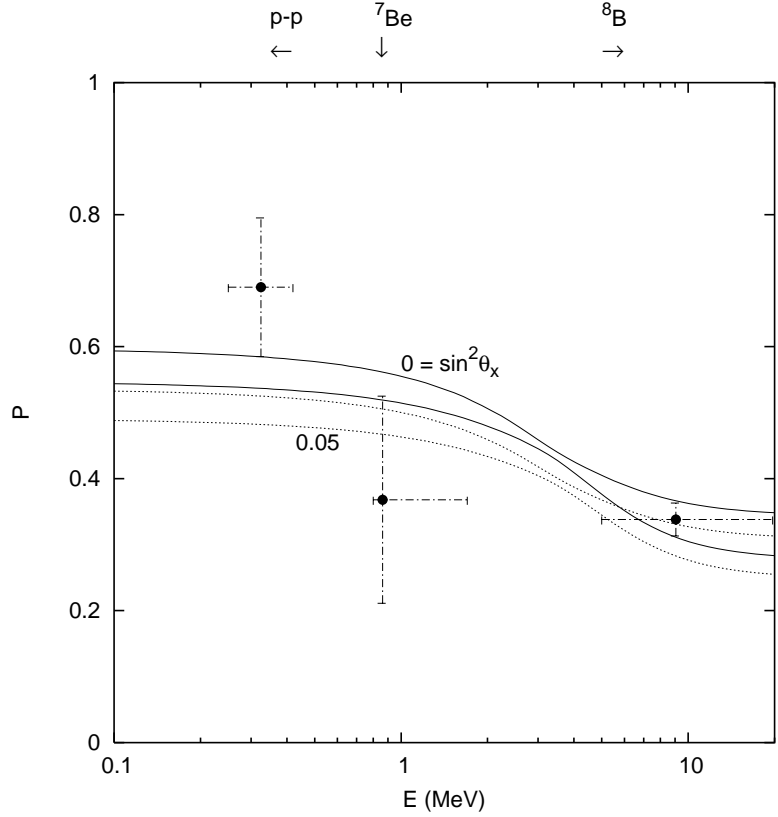}
\caption[]{LMA predictions for solar neutrino survival probability
versus neutrino energy for $\sin^2\theta_{x}=0$ (solid), 0.05 (dashed). 
The bands shows the range of predictions for $\delta m^2_{s}=(7.9\pm0.55)\times10^{-5}$~eV$^2$ and $\tan^2\theta_{s}=0.45\pm 0.06$. The
data points are taken from Eqs.~(\ref{eq:PH}, \ref{eq:PL}) and (\ref{eq:PI}),
the latter two of which were obtained using the SSM predictions 
for $\beta_L$ and $\beta_I$.
\label{fig:prob}}
\end{figure}

The LMA predictions for the solar neutrino survival probability versus
neutrino energy are shown in Fig.~\ref{fig:prob} for $\sin^2\theta_{x}
=0$ and 0.05 with $\delta m^2_{s}$ and $\theta_{s}$ varying over their
currently preferred ranges. Current data (taken from Eqs.~\ref{eq:PH},
\ref{eq:PL} and \ref{eq:PI}) are also shown, where the SSM constraints
in Eqs.~(\ref{eq:BL}) and (\ref{eq:BI}) have been used to determine
$P_L$ and $P_I$. The oscillation probability of the low-energy
(intermediate-energy) neutrinos is slightly higher (lower) than the
LMA prediction, although the discrepancies are currently only at the
$1\sigma$ level. Increasing $\theta_{x}$ lowers the LMA curve, which
improves the fit to the intermediate energy data, but at the expense
of a worse fit to the low energy data. Therefore reductions in the
uncertainties from future solar neutrino experiments will provide a
critical test of the LMA solution (see Sec.~3).

%We note that the values of $P_L$ and
%$P_I$ using the latest SSM deviate from the LMA predictions much
%less than those using the previous SSM. 
%Therefore in light of the updated SSM, the consistency of the LMA
%solution to the solar neutrino problem is improved.

Using the LMA expressions for $P_L$, $P_I$ and $P_H$, contours of
constant $\delta m^2_{s}$ and $\tan^2\theta_{s}$ are plotted in
Fig.~\ref{fig:PLMA2} in the $(P_H,P_L)$ and $(P_H,P_I)$ planes
for $\sin^2\theta_{x} = 0$ and 0.05. The
data points and larger error bars are taken from Eqs.~(\ref{eq:PH},
\ref{eq:PL}) and (\ref{eq:PI}), and the smaller error bars illustrate the
expected uncertainties 
from future measurements of $R_{pp}^{CC}$, $R_{pp}^{ES}$, $R_{Be}^{CC}$,
$R_{Be}^{ES}$, $R_{SNO}^{CC}$ and $R_{SNO}^{NC}$ with SSM constraints 
imposed (see the next section). The effect of increasing $\theta_x$ on
the compatability of the LMA solution with the data is  evident.

\section{Future data}

\subsection{Model independent analysis}

\subsubsection{Without SSM constraints}

Future experiments such as MOON~\cite{Doe:2003kq} or 
LENS~\cite{Lasserre:2002ke} or  will be able to provide better
measurements of CC scattering of the low and intermediate energy
neutrinos, {\it i.e.}, of
\begin{eqnarray}
R_{pp}^{CC} &=& \beta_L P_L \,,
\\
R_{Be}^{CC} &=& \beta_I P_I \,,
\end{eqnarray}
perhaps at the 2.5\% level~\cite{apsstudy}.  However, these measurements
will still not separate the flux normalization from the survival
probability. In order to determine the flux normalization from data
alone ({\it i.e.}, without imposing theoretical inputs from the SSM), a
process with a NC component must be used, such as the
elastic scattering (ES) measurements at BOREXino~\cite{Alimonti:2000xc}, CLEAN~\cite{McKinsey:2004rk}, 
HERON~\cite{Adams:1999yk} and KamLAND~\cite{kamsol}. They will measure
\begin{eqnarray}
R_{pp}^{ES} &=& \beta_L P_L(1-r_L) + \beta_L r_L \,,
\\
R_{Be}^{ES} &=& \beta_I P_I(1-r_I) + \beta_I r_I \,,
\end{eqnarray}
where $r_j$ is the ratio of the NC to $\nu_e$ cross-sections for
scattering off electrons in the appropriate energy regime. CLEAN plans
to measure $R_{pp}^{ES}$ to 1.1\% and $R_{Be}^{ES}$ to 2.6\%, HERON
plans to measure $R_{pp}^{ES}$ to 3.5\%, and BOREXino plans to measure
$R_{Be}^{ES}$ to 5\%~\cite{apsstudy}. KamLAND can measure $R_{Be}^{ES}$ to 
about 11\%~\cite{kamsol}. Once $R_{pp}^{ES}$ and $R_{Be}^{ES}$ are
measured, all six parameters (three flux normalizations and three
survival probabilities) will be determined by neutrino
data. Furthermore, SNO expects to reduce the uncertainties on its CC
and NC measurements to about 5.5\% and 6.4\%, respectively, in its
third phase~\cite{private}. 
When combined with its previous measurements, the SNO
uncertainties will be about 3.5\% and 4.8\%, respectively. The future
expectations for these uncertainties are summarized in
Table~\ref{tab:future}, where we list the smallest uncertainty in each channel.

\begin{table}[t]
\caption[]{Expected uncertainties from future solar neutrino measurements
of CC, NC and ES processes.
\label{tab:future}}
\centering\leavevmode
\begin{tabular}{c|ccc}
& $pp$ & $^7$Be & $^8$B\\
\hline
CC & 2.6\% & 2.5\% & 3.5\%\\
NC & $-$ & $-$ & 4.8\%\\
ES & 1.1\% & 2.6\% & $-$\\
\hline
\end{tabular}
\end{table}

The $pp$ measurements can be used to determine $\beta_L$ and $P_L$ via
\begin{eqnarray}
\beta_L &=& \left[ R_{pp}^{ES} - R_{pp}^{CC}(1-r_I)\right]/r_I \,,
\\
P_L &=& R_{pp}^{CC} r_I / \left[ R_{pp}^{ES} - R_{pp}^{CC}(1-r_I)\right] \,,
\end{eqnarray}
and similarly $\beta_I$ and $P_I$ can be determined from $R_{Be}^{ES}$
and $R_{Be}^{CC}$. Using the six measurements $R_{pp}^{CC}$,
$R_{Be}^{CC}$, $R_{SNO}^{CC}$, $R_{SNO}^{NC}$, $R_{pp}^{ES}$ and
$R_{Be}^{ES}$ with the projected uncertainties shown in
Table~\ref{tab:future}, the values of $\beta_L$, $\beta_I$ and
$\beta_H$ can be determined independently from any solar model
assumptions, with uncertainties of about 16\%, 22\% and 5\%,
respectively, assuming the central values of the $\beta_j P_j$ and
$\beta_H$ remain about the same (Eqs.~\ref{eq:BLPL}--\ref{eq:BH}) and that
the best-fit values of $\beta_L$ and $\beta_I$ are close to unity. The
constraints on $\beta_L$ and $\beta_I$ are not nearly as precise as
the $pp$ and $^7$Be measurements themselves because the NC component,
from which the value of $\beta$ is inferred, is suppressed by the
smaller NC cross-section for $\nu_\mu$ and $\nu_\tau$, compared to the
ES cross-section for $\nu_e$. The resulting uncertainties of $P_L$, $P_I$ and
$P_H$ would be about 14\%, 12\% and 6\%, respectively, if solar model
constraints are not imposed.

\subsubsection{With SSM constraints}

In the future, using the CC measurements of the $pp$ and $^7$Be
neutrinos to better determine $\beta_L P_L$ and $\beta_I P_I$,
respectively, the uncertainties of $P_L$ and $P_I$ will be reduced to
3\% and 11\%, respectively, when the SSM constraints are used; the
uncertainty of $P_I$ does not improve much since the SSM intermediate
energy flux is known to only 12\%. The large SSM uncertainty of
$\beta_H$ in Eq.~(\ref{eq:BH2}) does not provide any appreciable
reduction in the overall uncertainty of $\beta_H$ when combined with
SNO NC data.

\subsection{LMA analysis}

\subsubsection{Without SSM constraints}

Once the six measurements in Table~\ref{tab:future} are made, the six
parameters $\beta_L$, $\beta_I$, $\beta_H$, $\delta m^2_{s}$,
$\theta_{s}$ and $\theta_{x}$ will in principle be determined from
$R_{SNO}^{CC}$, $R_{SNO}^{NC}$, $R_{Be}^{CC}$, $R_{Be}^{ES}$,
$R_{pp}^{CC}$ and $R_{pp}^{ES}$, without needing any other inputs such
as SSM fluxes or KamLAND data. 

\begin{figure}[ht]
\centering\leavevmode
\includegraphics[width=3in]{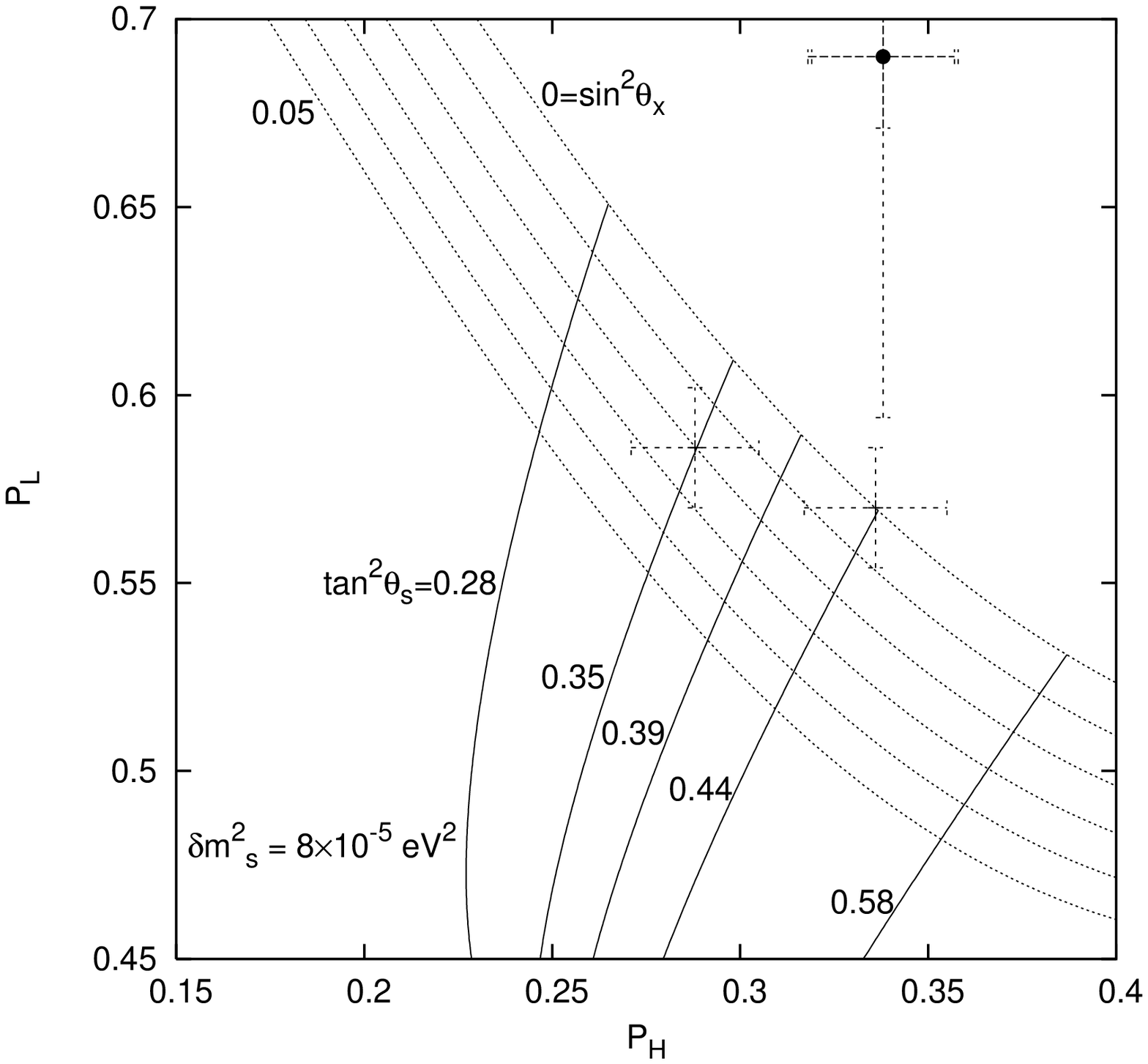}
\includegraphics[width=3in]{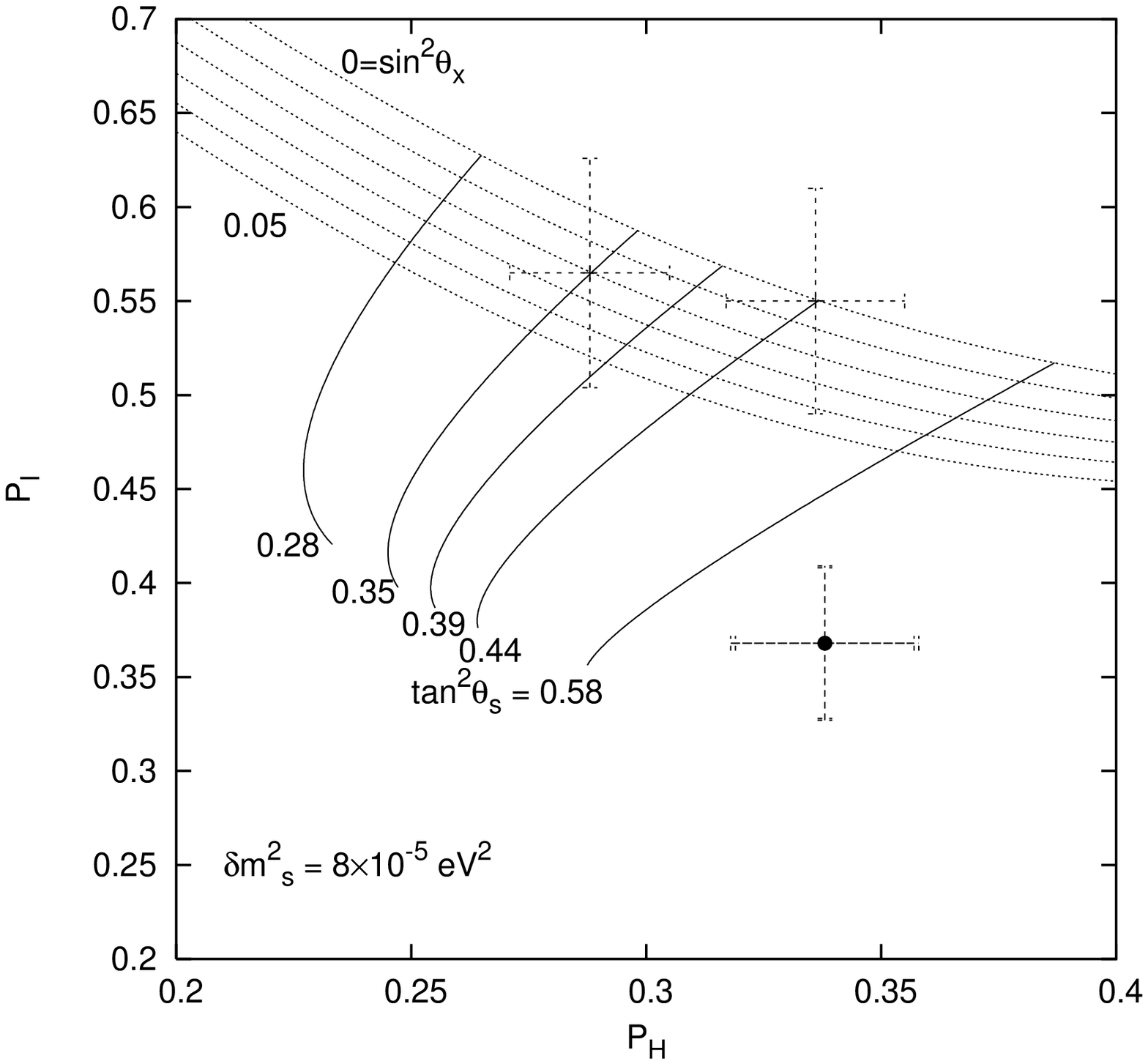}
\caption[]{Contours of $\theta_{s}$ and $\theta_{x}$ in the ($P_H,P_L$) 
and  ($P_H,P_I$) planes, for $\delta m^2_{s} = 8\times10^{-5}$~eV$^2$.
The bold data points are the current central values from Eqs.~(\ref{eq:PH},
\ref{eq:PL}) and (\ref{eq:PI}). The larger (smaller) error bars are the 
projected uncertainties
from the future measurements listed in Table~\ref{tab:future} without (with) 
SSM constraints imposed. While adding SSM constraints to future
data reduces the uncertainty in $P_L$ significantly, $P_I$ and $P_H$ are
essentially unimproved.
Two representative future measurements (with SSM flux constraints) 
are indicated at $(\tan^2 \theta_s, \sin^2 \theta_x)=(0.35,0.03)$ and (0.44,0).
 \label{fig:PLMA}}
\end{figure}

Figure~\ref{fig:PLMA} shows contours of constant $\theta_{s}$ and $\theta_{x}$ in two-dimensional subspaces of probability space for 
$\delta m^2_s=8\times 10^{-5}$ eV$^2$. The current best
fit values for these parameters are also shown in the figure.
The central value of $P_L$ is accommodated only 
for $\sin^2 \theta_x <0$, which is
 unphysical, while the central value of $P_I$ is accommodated for 
values of $\sin^2 \theta_x$ that violate the CHOOZ bound. However, there
is adequate consistency within error bars.
As noted earlier, if the flux normalizations
$\beta_L$ and $\beta_I$ of the SSM are correct, then the current
central values of $P_L$ and $P_I$ are slightly high and low, respectively. 
Future solar neutrino
measurements will be able to provide a more definitive statement about
this minor discrepancy, even without making any SSM assumptions. It is also
clear from Fig.~\ref{fig:PLMA} that a more precise {\it experimental}
determination of $\beta_I$ and especially $\beta_L$ would be needed to
provide significant improvements in the determination of oscillation
parameters if the predictions of the SSM are not used.

\subsubsection{With SSM constraints}

\begin{figure}[ht]
\centering\leavevmode
\includegraphics[width=3in]{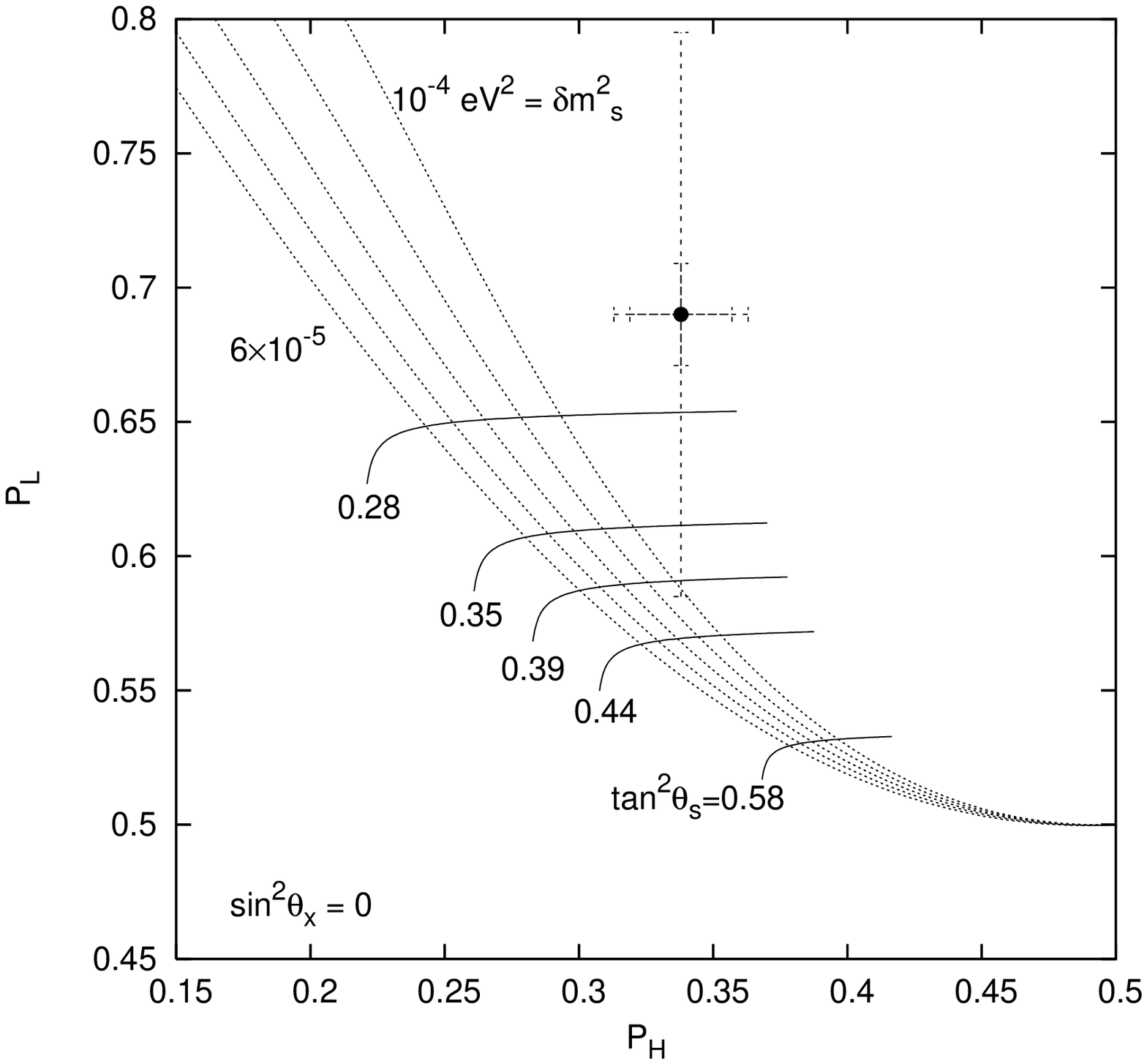}
\includegraphics[width=3in]{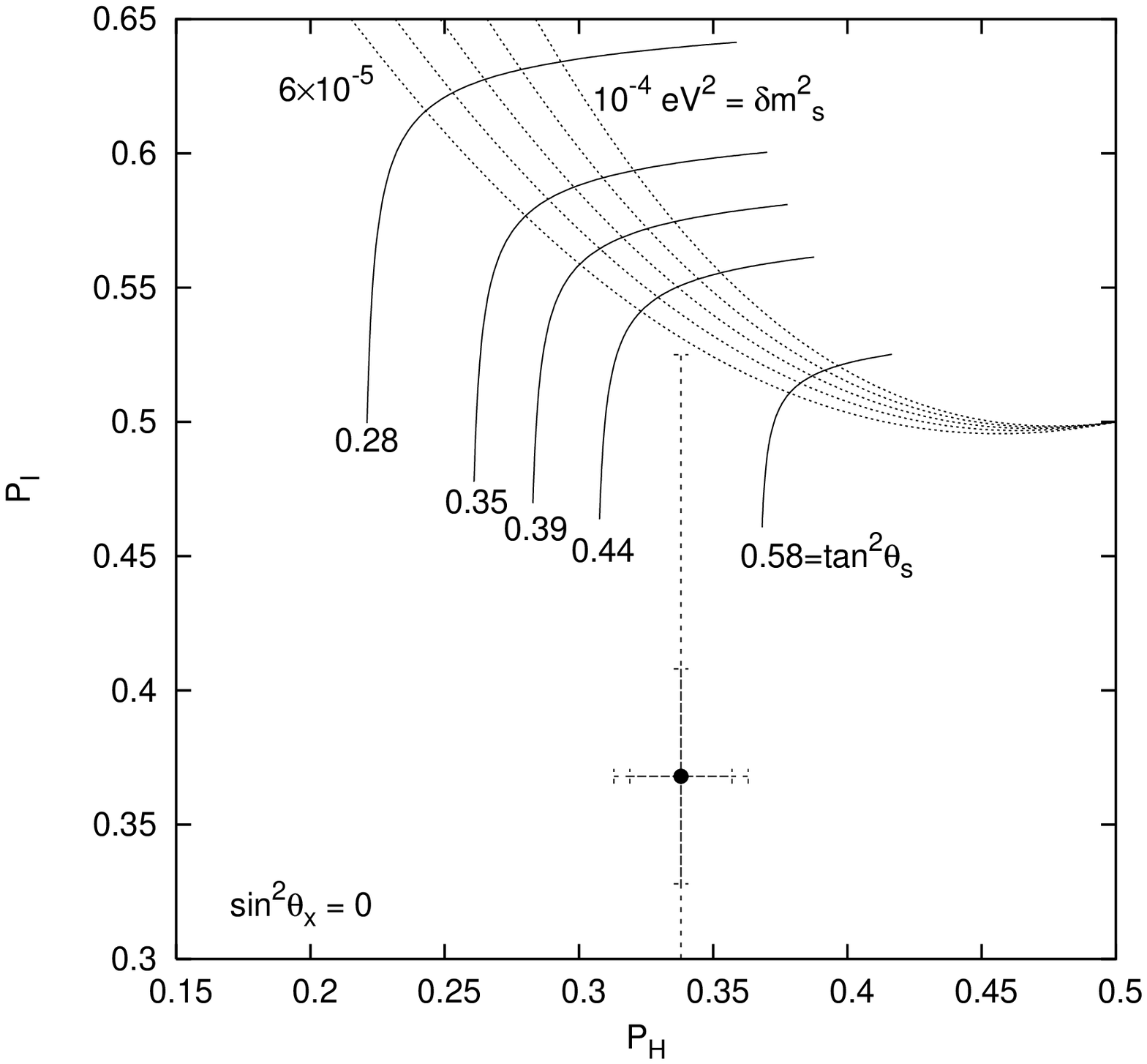}\\
\includegraphics[width=3in]{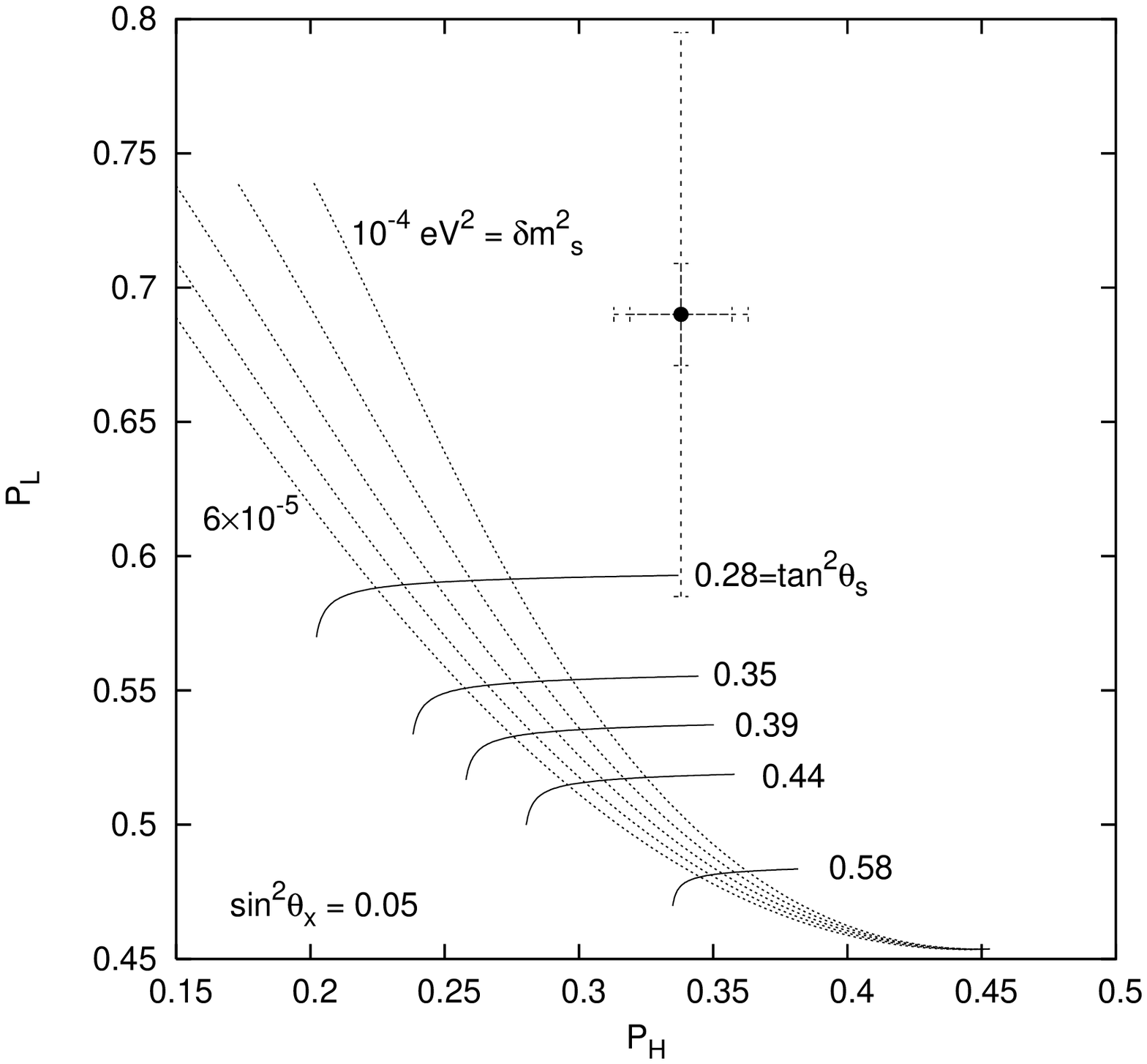}
\includegraphics[width=3in]{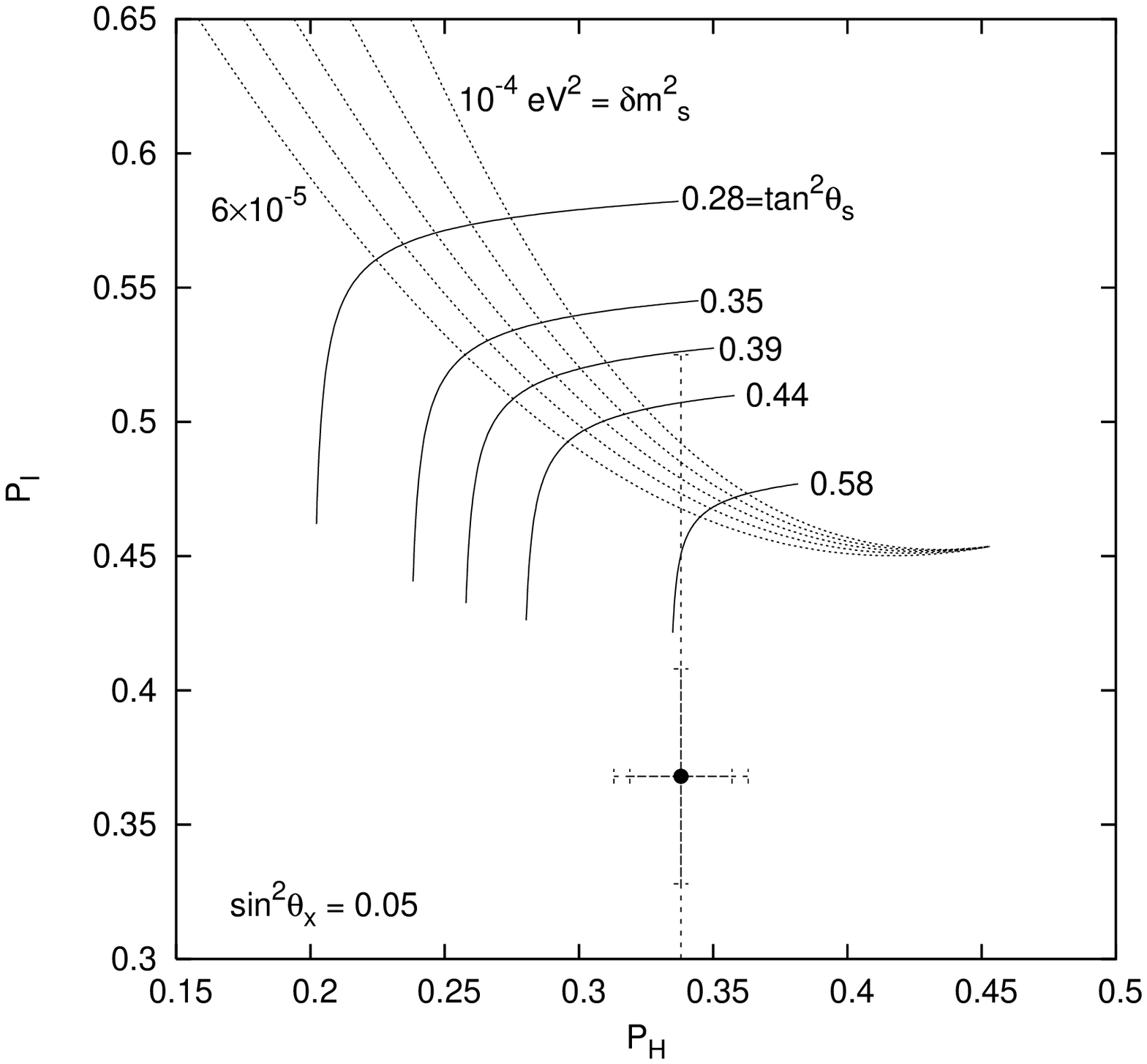}
\caption[]{
Contours of $\theta_{s}$ and $\delta m^2_{s}$ in the
($P_H,P_L$) and ($P_H,P_I$) planes for $\sin^2\theta_{x}=0$ and 0.05.
The bold data point and larger error bars are taken from Eqs.~(\ref{eq:PH},
\ref{eq:PL}) and (\ref{eq:PI}), the latter two of which were obtained 
using SSM constraints.
The smaller error bars are projected improvements in the uncertainties
from future measurements of $R_{pp}^{CC}$, $R_{pp}^{ES}$, $R_{Be}^{CC}$,
$R_{Be}^{ES}$, $R_{SNO}^{CC}$ and $R_{SNO}^{NC}$ with SSM constraints 
imposed. \label{fig:PLMA2}}
\end{figure}

In the future, after $R_{pp}^{CC}$ is measured and the SSM
constraint (Eq.~\ref{eq:BL}) is imposed, $P_L$ would be determined to
about 3\%. These projected uncertainties are shown as the smaller
error bars in Figs.~\ref{fig:PLMA2} and \ref{fig:PLMA}.

With the SSM constraints imposed, measurements of $\tan^2\theta_{s}$
and $\sin^2\theta_{x}$ can be made with $1\sigma$ uncertainties of order 0.03
and 0.015, respectively, by future solar neutrino experiments. As is
evident from Figs.~\ref{fig:PLMA2} and \ref{fig:PLMA}, to be consistent
with the LMA solution, 
the current central values of $P_L$ and
$P_I$ should shift when the new measurements are made.

We note that if $R_{Cl}$ were about 1$\sigma$ higher, then both
$\beta_L P_L$ and $\beta_I P_I$ would be about at the values predicted
by the best-fit LMA parameters and the SSM; they are coupled together
since $\beta_L P_L$ cannot be determined from $R_{Ga}$ without knowing
$\beta_I P_I$, which is determined from $R_{Cl}$ and $\beta_H P_H =
R_{SNO}^{CC}$. Since the primary constraint on $\beta_I P_I$ comes
from $R_{Cl}$, to which $\beta_I P_I$ only makes a 20\% contribution,
the current determination of $\beta_I P_I$ has a large
uncertainty. Future measurements of $\beta_I P_I$ in $R_{Be}^{CC}$
will not only significantly reduce the uncertainty on $\beta_I P_I$,
but also will test the Chlorine measurement.

\section{Summary}

We have shown that current neutrino data can determine the survival
probability of only the high-energy neutrinos if SSM flux
normalizations are not used, and that if the SSM flux constraints are
imposed, the LMA solution provides a consistent explanation of all
solar neutrino data, although minor discrepancies remain at the
$1\sigma$ level. In particular, the implied flux of low (intermediate)
energy neutrinos appears to be slightly high (low), if the LMA
solution is correct. Including the small mixing effects of the third
neutrino cannot compensate for both inconsistencies simultaneously.

Future measurements of low and intermediate energy neutrinos will
provide a much stricter test of the viability of the LMA solution 
independently of SSM predictions for the fluxes. See Table~\ref{tab:summary} 
for a summary of our results. If reliance is placed in the
SSM $pp$ flux normalization, the test becomes conclusive. If the LMA solution
survives without modification, it may be possible to demonstrate that 
$\theta_x$ is nonzero at more than $2\sigma$.

\begin{table}[t]
\centering\leavevmode
\begin{tabular}{c|c|cccccc}
Data & SSM imposed? & $\delta(\beta_L)$ & $\delta(\beta_I)$ &$\delta(\beta_H)$ &$\delta(P_L)$ &$\delta(P_I)$ &$\delta(P_H)$  \\
\hline
Current & No & $-$ & $-$ & 6 & $-$ & $-$ & 8 \\
Current & Yes& 1 & 12 & 6 & 15 & 43 & 8 \\
\hline
Future & No & 16 & 22 & 5 & 14 & 12 & 6 \\
Future & Yes & 1 & 11 & 5 & 3 & 11 & 6 \\
\hline
\end{tabular}
\caption[]{Percentage uncertainties in the flux normalizations and survival 
probabilities from current and future solar neutrino measurements. 
From the current data, $\beta_L$ and $\beta_I$ (and hence $P_L$
and $P_I$) cannot be determined without SSM input.
For the
future data we have assumed that the central values of $\beta_j P_j$ and 
$\beta_H$ coincide with those of the current measurements 
(see Eqs.~\ref{eq:BLPL}--\ref{eq:BH}) and that the best-fit
values of $\beta_L$ and $\beta_I$ are close to unity. 
\label{tab:summary}}
\end{table}

\vspace{0.2in}

\section{Acknowledgments}

We thank J.~Learned for a stimulating discussion. VB and DM
 thank the Aspen Center for Physics for hospitality.  
This research was supported by the U.S. Department of Energy under
Grants No. DE-FG02-95ER40896 and DE-FG02-01ER41155, by the NSF under 
Grant No. EPS-0236913, by the State of Kansas through Kansas Technology 
Enterprise Corporation and by the
Wisconsin Alumni Research Foundation.

\end{document}